\documentclass[final]{elsart5p}

 \usepackage{graphicx}

\usepackage{amssymb,latexsym}
\usepackage{wasysym}
\usepackage{textcomp}

\begin{document}

\begin{frontmatter}

\title{Coherent imaging of extended objects}
\author{E. Brainis \corauthref{EB}},
\ead{e.brainis1@physics.ox.ac.uk}
\author{C. Muldoon},
\author{L. Brandt},
\author{A. Kuhn}
\address{Department of Physics, University of Oxford, Parks Road OX1~3PU Oxford, United Kingdom }
\corauth[EB]{Corresponding author.}


\begin{abstract}
When used with coherent light, optical imaging systems, even diffraction-limited, are
inherently unable to reproduce both the amplitude and the phase of a two-dimensional field distribution
because their impulse response function varies slowly from point to point (a property 
known as non-isoplanatism). For sufficiently small objects, this usually results in a
phase distortion and has no impact on the measured intensity. Here, we show that the intensity distribution can 
also be dramatically distorted when objects of large extension or of special shapes are imaged. We
illustrate the problem using two simple examples: the pinhole camera and the 
aberration-free thin lens. The effects predicted by our theorical analysis are also confirmed by experimental 
observations. 
\end{abstract}

\end{frontmatter}

\section{Introduction}
Current technology, especially the ability to
manufacture aspherical surfaces, allows lenses and mirrors to be
designed that minimize the most important geometrical aberrations.
Such optical elements are nearly ideal instruments obeying the laws
of Gaussian optics even for far off-axis points and non-paraxial
rays. Self-luminous objects or objects illuminated with
\emph{incoherent} light can be imaged with high fidelity. The
quality of the resulting image is merely fixed by the resolution of
the instrument which is related to its numerical aperture. The
instrument itself is said to be \emph{diffraction-limited} and can
be considered as a linear filter for the intensity of light
\cite{BW,Good}.

When imaging objects with \emph{coherent} light, the conditions for
accurate image formation are more severe since both
the relative amplitudes and the relative phases of the object points 
have to be mapped to the corresponding image points
 (up to the resolution capability of the
instrument). This only happens if the response of the optical instrument
to the field from a given point source is independent of 
its position in the object plane, or in other words, if the coherent impulse
 response \cite{BW,Good} of the system is \emph{space-independent}. The
instrument then acts as a \emph{linear filter} for the complex field amplitude. 
According to the terminology of
\cite{BW}, such an instrument is said to be \emph{isoplanatic}.
In general, even aberration-free optical instruments designed to
map some planar object to an image plane under
incoherent illumination do not meet this condition. Some spatial phase distortion is
unavoidably introduced, which, when combined with a
finite resolution, severely modifies the intensity distribution of the
image. This was first recognized by Dumontet  
\cite{Dumontet}, and later by Tichenor and Goodman \cite{TG},
who showed that a thin lens can only be considered
as an isoplanatic system if both the object and the
image lie on spherical surfaces $\mathcal{S}_o$ and
$\mathcal{S}_i$, tangent to the geometrical-optics
object and image planes $\mathcal{O}$ and $\mathcal{I}$ respectively, and
having their center of curvature in the plane of the lens (see
Sec.~\ref{sec3}).

In practice, an aberration-free optical instrument can be
treated as an ideal coherent-light imaging system
whenever the spherical surfaces $\mathcal{S}_o$ and $\mathcal{S}_i$
can be approximated by their tangent planes $\mathcal{O}$ and
$\mathcal{I}$. We emphasis that this is only
viable when the object to be imaged is very small and
lies close to the optical axis. A weaker imaging condition has been 
obtained by Tichenor and Goodman who showed that non-isoplanatism 
of a thin lens has a negligeable
effect on the \emph{intensity distribution} of the image if
the object diameter is smaller than about a quarter of the
lens diameter \cite{TG}. In this paper, we investigate
the effect of non-isoplanatism on coherent image
formation when this condition is not met. 
That situation may be encountered in many
fields of optics where large-sized objects are
imaged through powerful limited-aperture instruments, as in coherent
far-field microscopy, optical
lithography \cite{ito}, holographic data-storage \cite{heanue}, and dipole-trapping of neutral atoms
\cite{Brandt}. In particular arrays of coherently emitting point sources, like individual trapped atoms excited by the same laser beam \cite{Meschede,Miroshnychenko}, will be subject to the phenomena discussed here. Effects similar to those that we described in the context of imaging are also expected when lenses are used to perform the spatial Fourier transform of a two-dimensionnal field distribution, as or holographic dipole-trapping of atoms \cite{Bergamini}.

\section{Non-isoplanatism of a pinhole camera}\label{sec2}
\begin{figure}[b]
\centerline{\includegraphics*[width=8cm]{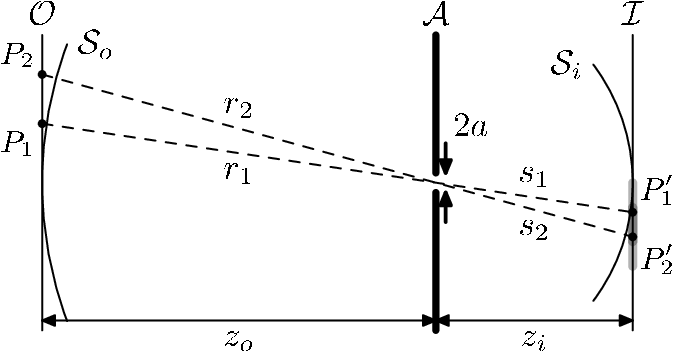}}
\caption{Scheme for coherent image formation by a pinhole camera. The points
$P'_1$ and $P'_2$ in the image plane $\mathcal{I}$ are the
geometrical images of the point objects $P_1$ and $P_2$ in the
object plane $\mathcal{O}$. The light-gray zones around $P'_1$
and $P'_2$ in the
$\mathcal{I}$-plane represent the individual
diffraction patterns resulting from illumination of
the pinhole by $P_1$ and $P_2$ respectively. The dark-gray zone
represents the area where the individual
diffraction patterns overlap and interfere.} \label{phc}
\end{figure}

The problem of non-isoplanatism can be best
understood by considering the simple example of coherent image
formation through a pinhole camera (Fig.~\ref{phc}).

\begin{figure}[t]
\centerline{\includegraphics*[width=8cm]{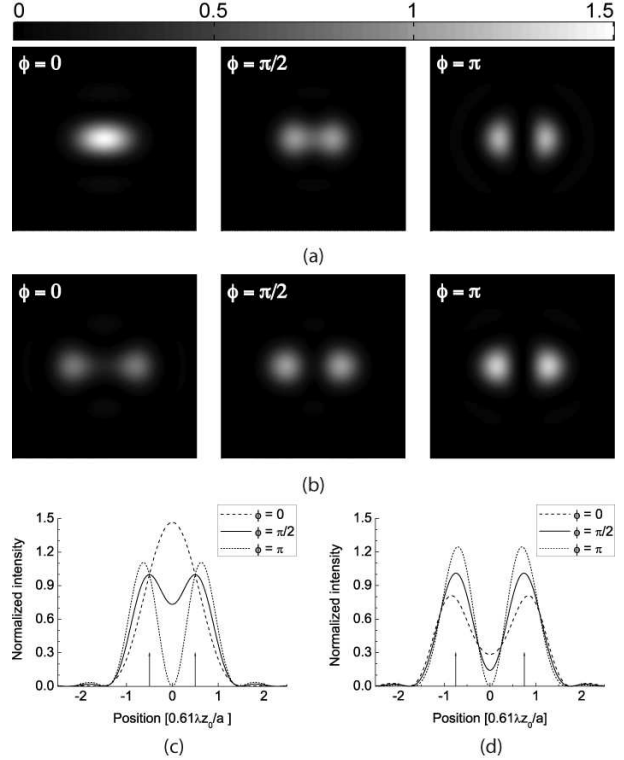}}
\caption{Intensity profile in the image plane of a
diffraction-limited pinhole camera due to the interference of the
Airy patterns of two coherent point sources. The distance
$|x_{o,1}-x_{o,2}|$ between the sources is equal to (a) $0.61
\lambda z_{o}/a$ (the Rayleigh resolution limit), and (b) $0.915 \lambda
z_{o}/a$  (1.5 times the Rayleigh resolution limit). The Airy
patterns have been normalized so that their peak intensity is 1. Panel (c) and (d) show the intensity profiles along the $x$-axis corresponding to the images in panel (a) and (b), respectively. Vertical arrows in panels (c) and (d) point the positions of the centers of the interfering Airy patterns.}\label{imagePHC}
\end{figure}

Consider the following situation: Two mutually coherent
point-objects, $P_1$ and $P_2$, are imaged through a small pinhole (of radius $a$)
from an object plane $\mathcal{O}$ to an image plane $\mathcal{I}$.
We call
$(x_{o,k},y_{o,k})$ the transverse coordinates of the point sources
$P_k$ ($k\in\{1,2\}$), and $(x_{i,k},y_{i,k})$ the
transverse coordinates of their geometrical images $P'_k$. For simplicity, we
chose $y_{o,k}=y_{i,k}=0$ and assume that the point sources emit
 in phase. We assume that the pinhole is so small that the
resolution is limited by diffraction.
The ``images"
of $P_1$ and $P_2$ are two Airy patterns centered on the geometrical
image points $P'_1$ and $P'_2$. Let's consider that $P'_1$ and $P'_2$ 
are at the resolution limit
according to the Rayleigh criterion $x_{i,1}-x_{i,2}=0.61 \lambda
z_{i}/a$. Due to the 
geometry displayed in Fig.~\ref{phc}, the relative phase
 of their respective Airy patterns is $\phi=2\pi (r_{1}-r_{2})/ \lambda$.  This shows that
 \emph{the phase relation between points is not mapped properly from the object
 plane to the image plane}. Furthermore, \emph{the relative phase $\phi$ varies as a function
 of  $x_{o,m}\equiv(x_{o,1}+x_{o,2})/2$, the mean distance of the point sources to the optical
 axis}. The phase $\phi$ is obviously null if
$x_{m}=0$, but it already reaches the value $\pi/2$ when
$|x_{o,m}|=a/2.44$. Since $a$ is usually about 1~mm for a pinhole camera, $\phi$ varies
 extremely rapidly when the two point sources are moved over the object plane. 
Fig.~\ref{imagePHC} shows the intensity distribution in the image
 plane resulting from the interference of two Airy patterns with relative phase $\phi=0$, $\pi/2$, and
$\pi$. In panel (a) the distance between the sources corresponds to
the Rayleigh resolution criterion, as discussed before; in panel (b)
this distance is increased by a factor of 1.5. Panel (c) and (d) display 
the intensity profile along the $x$-axis for images in panel (a) and (b), respectively.
 Note that the intensity distribution for $\phi=\pi/2$ is the same as for incoherent
 sources. Because of this
interference, the image points may be unresolved even
if their relative positions fulfill the Rayleigh criterion.

The preceding example shows that different intensity distributions must be expected from
identical object patterns depending
on their positions in the object plane. This is due to an incorrect phase mapping between 
the object and image plane. The relative phase of an object point and its image depends
 on the position of the point source in the object plane. We refer to this situation as
 the \emph{non-isoplanatism} of coherent imaging. Considering the pinhole camera as a linear
 optical system \cite{Good}, non-isoplanatism is related to the space-variance of its
 impulse response. We develop this point of view hereafter.

Since the pinhole camera is a linear system, the complex field
amplitudes in the object and image planes, $U_o(x_o,y_o)$ and $U_i(x_i,y_i)$ satisfy the 
integral relation
\begin{displaymath}
U_i(x_i,y_i)=\iint h(x_{i},y_{i}|x_{o},y_{o}) \ U_o(x_o,y_o) \d x_o \d y_o.
\end{displaymath} 
The impulse response of the camera
(in the paraxial and far-field approximation) is
given by the Fraunhofer integral
\begin{eqnarray}\label{hPLC}
h(x_{i},y_{i}|x_{o},y_{o})&=&\frac{1}{\lambda^2}\  \frac{\e^{i 2 \pi (r+s) / \lambda}}{r s} \iint_{\mathcal{A}} \e^{-i 2 \pi (f_{x}\xi+f_{y}\eta)} \d\xi \d\eta, \nonumber \\
&=&\frac{1}{\lambda^2}  \ \frac{\e^{i 2 \pi (r+s) / \lambda}}{r s} \
\delta_{a}(f_{x},f_{y}),
\end{eqnarray}
where $(x_o,y_o)$ are the coordinates of the point
source in the object plane, $(x_i,y_i)$ are the coordinates of the
``observation point'' in the image plane. The integration domain
$\mathcal{A}$ is the pinhole opening disc, while
\begin{eqnarray*}
r&=&\sqrt{x_{o}^{2}+y_{o}^{2}+z_{o}^{2}},\\
s&=&\sqrt{x_{i}^{2}+y_{i}^{2}+z_{i}^{2}}
\end{eqnarray*}
are the distances from the point source and the observation point to
the pinhole center, and
\begin{eqnarray*}
f_{x}&=&\frac{1}{\lambda}\left(\frac{x_{i}}{s}+\frac{x_{o}}{r}\right), \\
f_{y}&=&\frac{1}{\lambda}\left(\frac{y_{i}}{s}+\frac{y_{o}}{r}\right)
\end{eqnarray*}
 are the spatial frequencies of the plane waves diffracted by the pinhole%
.  On the second line of (\ref{hPLC}), the function
 \begin{equation}
 \delta_{N}(x,y)=|N| \  \frac{J_{1}(2\pi \ N \ \sqrt{x^{2}+y^{2}})}{\sqrt{x^{2}+y^{2}}} \quad (N\in \mathbb{R}_{0})
 \end{equation}
 has been introduced ($J_{1}(x)$ is a Bessel function of first kind). Its square $\delta^{2}_{N}(x,y)$ is
 the usual intensity point spread function or Airy pattern of diffraction-limited optical systems and
 it is normalized so that $\lim_{|N|\rightarrow \infty} \delta_{N}(x,y)=\delta(x,y)$ (the Dirac distribution).
  The ``scaling'' property,
  $\delta_{N}(C x, C y)=\delta_{CN}(x, y)/C^{2}$ for any $C\in \mathbb{R}_{0}$,
  can be used to simplify (\ref{hPLC}):
 \begin{eqnarray}\label{HPLC}
 h(x_{i},y_{i}|x_{o},y_{o})&=&|M| \ \e^{i \frac{2 \pi}{\lambda} (r+s) } \times \nonumber \\
 &\phantom{=}&  \delta_{\frac{a}{\lambda s}}(x_{i}-M x_{o},y_{i}-My_{o}),
 \end{eqnarray}
where $M=-s/r$ is the geometric
\emph{magnification ratio} of the camera. Note that we can make the 
approximation $M\approx-z_{i}/z_{o}$ because $r$ and $s$ vary only very slightly over the object and image planes.
For the same reason,  $\delta_{a/(\lambda s)}(x,y)\approx
\delta_{a/(\lambda z_i)}(x,y)$.

The phase factor in Eq.~(\ref{HPLC}) plays an important role. If the phase
factor were not there, the impulse function
corresponding to any point $(x_o,y_o)$ lying in the object plane
would be the same as the impulse function of the origin
$(0,0)$, but translated to the geometrical image point $(M
x_o,M y_o)$; in that case, the system would
be space-invariant, or \emph{isoplanatic}, and the
image of a coherent object with field amplitude $U_o(x_o,y_o)$ could
be computed by simply
convoluting $U_o(x_o,y_o)$ with the impulse function. In other
words, the pinhole camera would act as a linear filter that
reproduces the objet $U_o(x_o,y_o)$ with a lower resolution. The
example of Fig.~\ref{imagePHC} clearly shows that this is not the
case. Because of the phase factor, the impulse response function is different
for different points in the object plane.
This creates interference patterns that modify the image
more significantly than a simple blur. The image of a field profile
$U_o(x_o,y_o)$ is given by
\begin{eqnarray}\label{transPHCni}
U_i(x_i,y_i) &=&\frac{1}{|M|}\ \e^{i \frac{2 \pi}{\lambda} s } \iint
\ U_o(x_o,y_o) \  \e^{i \frac{2 \pi}{\lambda} r } \nonumber \\ &&
\delta_{\frac{a}{\lambda z_o}}(\frac{x_{i}}{M}-
x_{o},\frac{y_{i}}{M}-y_{o}) \ \d x_o
\ \d y_o.
\end{eqnarray}
To establish Eq.~(\ref{transPHCni}), we make use of
the fact that
\begin{displaymath}
\delta_{\frac{a}{\lambda z_i}}(x_{i}-M
x_{o},y_{i}-My_{o})=\frac{1}{M^2}\delta_{\frac{a}{\lambda
z_o}}(\frac{x_{i}}{M}- x_{o},\frac{y_{i}}{M}-y_{o})
\end{displaymath}
can be considered either as a function of $(x_i,y_i)$, the usual Airy
pattern centered on the geometrical image point $(Mx_o,My_o)$, or as
a function of $(x_o,y_o)$ centered on the object point
$(x_i/M,y_i/M)$. Note that in the first interpretation, the
first-zero full-width of the Airy pattern $e_i=1.22 \lambda z_i/a$
represents the region of the image plane most influenced by the
field originating from the point source at $(x_o,y_o)$ in the object
plane. In the second interpretation, the first-zero full-width
of the Airy pattern $e_o=1.22 \lambda z_o/a$ represents the area of
the object plane that most contributes to the field at the
observation point $(x_i,y_i)$ in the image plane. In the \emph{non-isoplanatic} situation described by Eq.~(\ref{transPHCni}),
important interferences occur when the phase factor $\exp{[i \frac{2
\pi}{\lambda} r ]}$ cannot be considered as constant over the
circular region of area $\pi (e_o/2)^2$ centered on $(x_i/M,y_i/M)$
in the object plane.

It is interesting to note that isoplanatism is recovered when one
considers the imaging problem from $\mathcal{S}_o$ to
$\mathcal{S}_i$ (see Fig.~\ref{phc}), where $\mathcal{S}_o$
($\mathcal{S}_i$) is a spherical surface of radius $z_o$ ($z_i$)
centered on the pinhole, and having its vertex on $\mathcal{O}$
($\mathcal{I}$). The spherical field distributions on
$\mathcal{S}_o$ and $\mathcal{S}_i$ are described
by the amplitudes $U_{S_o}(x_o,y_o)$ and $U_{S_i}(x_i,y_i)$,
respectively ($z$-coordinates are dependent
variables). These are related to the field amplitudes on
$\mathcal{O}$ and $\mathcal{I}$ by a phase relation:
$U_{S_o}(x_o,y_o)=U_{o}(x_o,y_o) \exp[i 2 \pi (r-z_o) /\lambda]$ and
$U_{S_i}(x_i,y_i)=U_{i}(x_i,y_i) \exp[-i 2 \pi (s-z_i) /\lambda]$.
Inserting these relations into
Eq.~(\ref{transPHCni}) and removing the constant
phase factor $\exp[i 2 \pi (z_o+z_i) /\lambda]$ one gets:
\begin{eqnarray}\label{transPHCi}
U_{S_i}(x_i,y_i)&=&\frac{1}{|M|} \ \iint \ U_{S_o}(x_o,y_o) \nonumber \\
&&\delta_{\frac{a}{\lambda z_o}}(\frac{x_{i}}{M}-
x_{o},\frac{y_{i}}{M}-y_{o}) \ \d x_o \ \d y_o
\end{eqnarray}
which is a convolution relation, as expected for an
isoplanatic imaging system. The physical reason why isoplanatism is
recovered when the object lies on the spherical
surface $\mathcal{S}_o$ is simple to understand: since all the point
sources are at same distance $z_o$ from the pinhole, their point
spread functions always interfere constructively. Note that there is
no real need to measure the image on $\mathcal{S}_i$ for it only
differs from the image in the plane $\mathcal{I}$ by a space-dependant phase
that any standard intensity detector is insensitive to.

\section{Non-isoplanatism of a thin lens}
\label{sec3}
The scenario exhibited in the simple example of the pinhole camera
actually occurs in all optical imaging systems. When imaging with
lenses or mirrors, however, the diffraction effects are less
dramatic than with a pinhole camera. The problem of phase
distortion remains though, and may sometimes induce
unwanted intensity modulation in the image.
\begin{figure}[h]
\centerline{\includegraphics*[width=8cm]{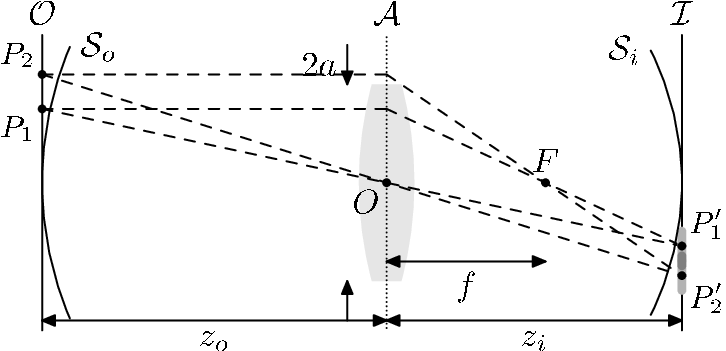}}
\caption{Scheme of coherent image formation by an aberration-free thin lens.
The points $P'_1$ and $P'_2$ in the image plane $\mathcal{I}$ are
the geometrical images of the point objects $P_1$ and $P_2$ in the
object plane $\mathcal{O}$. The light-gray zones
around $P'_1$ and $P'_2$ in the $\mathcal{I}$-plane represent the
individual diffraction patterns resulting from illumination of the
pinhole by $P_1$ and $P_2$ respectively. The dark-gray zone
represents the area where the individual diffraction patterns
overlap and interfere.} \label{tl}
\end{figure}

An aberration-free thin lens is an ideal diffraction-limited optical
system that acts locally on the impinging field as a pure phase
transparency $T(x,y)=\exp[-i\pi (x^2+y^2)/(\lambda f)]$, where $f$
is the focal length of the lens. The impulse response function of a
thin lens \cite{TG, Good} is an important concept that has proven to be
very useful for the design of optical systems and optical data
processing. It turns out that it is given by the same formula as for
the pinhole camera --- Eq.~(\ref{HPLC})
--- with the understanding that $z_{i}$ is now related to $z_{o}$
through the lens law $1/z_{o}+1/z_{i}=1/f$. To stress the analogy
between the pinhole camera and the thin lens systems, we have drawn
on Fig.~\ref{tl} the same information as on Fig.~\ref{phc}, and used
the same notation. In particular, note that $2a$ now represents the
diameter of the lens, which is usually considerably
larger than the aperture of the pinhole camera. As a consequence, the
diffraction effects will be weaker, and non-isoplanatism will be
less pronounced. Apart from this comment, the results and discussion
of the last section also apply to the thin lens
system. The relation between the field distribution
in the plane $\mathcal{O}$ and its image on the plane
$\mathcal{I}$ is non-isoplanatic and is given by
Eq.~(\ref{transPHCni}). As with the pinhole camera,
 isoplanatism can be recovered
--- Eq.~(\ref{transPHCi}) --- when the thin lens is used to image the spherical surface 
 $\mathcal{S}_{o}$ to $\mathcal{S}_{i}$. However, this is much less obvious here
 because, in contrast with the pinhole camera, a lens does not have an infinite field of view. That
$\mathcal{S}_{i}$ is the image surface of  $\mathcal{S}_{o}$ is therefore questionable.
This point requires a more carefull argumentation that we postpone to Sec.~\ref{sec4}.

Whether non-isoplanatism leads to interference
when a plane emitter is imaged to a plane receptor depends not only
on the size of the lens, but also on the size of the object and its
position with respect to the optical axis. The condition for
avoiding any interferences due to non-isoplanatism is that the phase
$2\pi r/\lambda$ in Eq.~(\ref{transPHCni}) varies by less than $\pi$
when the point corresponding to coordinates
$(x_o,y_o)$ explores the Airy pattern $\delta_{\frac{a}{\lambda
z_o}}(x_{i}/M- x_{o},y_{i}/M-y_{o})$ in the object plane. If this is
the case, the phase factor $\exp[i2\pi r/\lambda]$ can
be taken out of the integral in
Eq.~(\ref{transPHCni}). Since the fastest phase variation occurs
when the point moves radially off axis, the following condition is
obtained:
\begin{equation}\label{crit}
e_o\left(\rho^g_o+\frac{e_o}{4}\right)\ll \lambda z_o,
\end{equation}
where $\rho^g_o=\sqrt{(x_i/M)^2+(y_i/M)^2}$ is the
off-axis distance of the geometrical object point under
consideration and $e_{o}=1.22 \lambda z_{o}/a$ is the first-zero full width of the Airy pattern in the object plane. Note that Tichenor and Goodman
\cite{TG} found a similar condition following a different reasoning.
The main conclusions are:
\begin{itemize}
\item When imaging points very close to the optical axis ($\rho^g_o\ll e_o/4$),
interference due to non-isoplanatism does not occur if $a^2/\lambda
z_o\gg 1$. This last condition is always satisfied
in practice with lenses. (For a diffraction-limited pinhole camera
it fails to be satisfied.)
\item When imaging off-axis points ($\rho^g_o\gg e_o/4$), interference due to non-isoplanatism does not occur
if $\rho^g_o\ll a$, i.e. if the object points are not as far
off-axis as the edges of the lens.
\end{itemize}
When criterion (\ref{crit}) is satisfied for any point
on the object, Eq.~(\ref{transPHCni}) can be
written as
\begin{eqnarray}\label{quasi}
U_{i}(x_i,y_i)&=&\frac{1}{|M|} \e^{i\pi \frac{x_i^2+y_i^2}{\lambda (z_i-f)}}\ \iint \ U_{o}(x_o,y_o) \nonumber \\
&&\delta_{\frac{a}{\lambda z_o}}(\frac{x_{i}}{M}-
x_{o},\frac{y_{i}}{M}-y_{o}) \ \d x_o \ \d y_o,
\end{eqnarray}
where the phase factor comes from the second order approximation of
$s(x_i,y_i)+r(x_i/M,y_i/M)$ and constant phases
have been removed. An intensity detector in the
plane $\mathcal{I}$ will record the same image as in the isoplanatic
case
--- Eq.~(\ref{transPHCi}). It should however be noted that the
impulse response function is still \emph{non}-isoplanatic because
the phase curvature has not been removed. It is important to keep
this in mind when a phase-sensitive detector (hologram) is used
and/or if further optical processing is needed.

\section{Non-isoplanatism and large field imaging}\label{sec5}

According to the discussion in Sec.~\ref{sec3},
non-isoplanatism has no effect on the intensity detected in the
image plane if  condition (\ref{crit}) is
satisfied. In that case, Eq.~(\ref{quasi}) can be used instead of
Eq.~(\ref{transPHCni}). In practical applications
of coherent imaging, the assumption is usually made \cite{Good} that
the field mapping from the object space to the image space \emph{is}
given by Eq.~(\ref{quasi}) for any lens of a given optical system.
This assumption is very convenient from a theoretical point of view
since it makes the analysis of optical systems much simpler by the
use of standard Fourier optics methods. From a
technical point of view, however, good quality lenses (aspheric lenses,
for instance) usually have a small diameter because of manufacturing
constrains, and care must therefore be taken when implementing optical
systems based on Eq.~(\ref{quasi}). In this section, we show that
the slight difference between Eqs.~(\ref{quasi}) and
(\ref{transPHCni}) may lead to strong effects if the object does
not fulfill condition (\ref{crit}).

Let's first point out what
is wrong with Eq.~(\ref{quasi}) from a physical point of view.
Consider an object field $U_o(x_o,y_o)$ that is varying slowly on
the length scale $e_{o}=1.22 \lambda z_{o}/a$ of the peaked function
 $\delta_{a/(\lambda
z_o)}(x_{i}/M-x_{o},y_{i}/M-y_{o})$. Then,  according to Eq.~(\ref{quasi}),
one can write
\begin{equation}\label{perf}
|U_i(x_i,y_i)|^2=\frac{1}{|M|^2}\left|U_o\left(\frac{x_i}{M},\frac{y_i}{M}\right)\right|^2.
\end{equation}
This equation implies that the energy is  conserved:
 $\iint
|U_i(x_i,y_i)|^2 \d x_i \d y_i = \iint |U_o(x_o,y_o)|^2 \d x_o \d
y_o$.
 However, independently of how slowly the field varies in space,
Eq.~(\ref{perf}) cannot hold for far off-axis points when the lens
has a limited aperture. Radiation from far off-axis points (like $P_2$ on
Fig.~\ref{tl}) is partially lost, and energy cannot be conserved.
For instance, in the case of a plane wave travelling along the optical axis,
radiation from the neighborhood of $P_2$ will not be transmitted at all.
We can therefore conclude that Eq.~(\ref{quasi}) does not properly
account for energy loss due to the limited aperture of the lens,
especially for light originating from far off-axis
points. Taking the phase factor $\exp(i2\pi
r/\lambda)$ out of the integral sign in Eq.~(\ref{transPHCni})
breaks down the energy balance.

\begin{figure}[t]
\centerline{\includegraphics*[width=8.5cm]{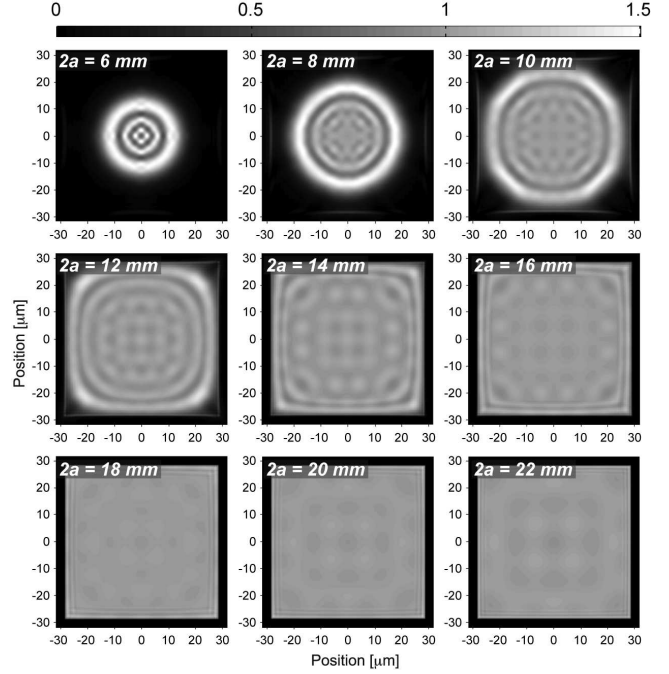}}
\caption{Intensity in the image plane $\mathcal{I}$ when a
large square aperture, centered on the optical axis, is imaged with a thin
lens. The square aperture is illuminated by a plane wave
travelling along the optical axis. The sides of the square are
$b=9.5$~mm. The different panels correspond to different lens
diameters. From left to right and top to bottom, the lens diameter
is progressively increased in 2~mm-steps from $2a=6$~mm to $2a=22$~mm.
The relevant parameters are: $\lambda=780$~nm, $f=12$~mm, and $z_o=2$~m. The scale of
the figures is expressed in microns. The intensity scale is such that 1
corresponds to the expected uniform intensity in the center
of the square in the limit of an infinite lens.} \label{iso1}
\end{figure}
To give a deeper insight of the effect of the phase
factor $\exp(i2\pi r/\lambda)$, we use Eq.~(\ref{transPHCni})
instead of Eq.~(\ref{quasi}) to compute the intensity distribution
in the image plane for a slowly varying object field. We now obtain
\begin{eqnarray*}
&&|U_i(x_i,y_i)|^2=\frac{1}{|M|^2}\left|U_o\left(\frac{x_i}{M},\frac{y_i}{M}\right)\right|^2 \nonumber\\
&& \times \left|\iint \e^{i \frac{\pi}{\lambda z_o}(x^2_o+y^2_o)} \
\delta_{\frac{a}{\lambda z_o}}(\frac{x_i}{M}-x_o,\frac{y_i}{M}-y_o)
 \ \d x_o \d y_o\right|^2
\end{eqnarray*}
instead of Eq.~(\ref{perf}). The paraxial approximation $r\approx
z_0+(x_o^2+y_o^2)/(2z_o)$
 has been used. Making a change of integration variables from $(x_{o},y_{o})$ to $(\xi,\eta)=(x_o,y_o)-(x_i,y_i)/M$, the preceding
equation can be written as
\begin{eqnarray}\label{perf2}
&&|U_i(x_i,y_i)|^2=\frac{1}{|M|^2}\left|U_o\left(\frac{x_i}{M},\frac{y_i}{M}\right)\right|^2 \nonumber\\
&& \times \left|\iint \e^{i \frac{2\pi}{\lambda z_o}(
\frac{x_i}{M}\xi+
\frac{y_i}{M}\eta+\frac{\xi^2}{2}+\frac{\eta^2}{2})} \
\delta_{\frac{a}{\lambda z_o}}(\xi,\eta)
 \ \d \xi \d \eta\right|^2.
\end{eqnarray}
The integral in Eq.~(\ref{perf2}) has the form of a \emph{Fresnel
diffraction integral}. For points $(x_i,y_i)$ which are sufficiently far away from the optical axis,
the quadratic phase terms $\xi^2/2$ and $\eta^2/2$ can, as a first order approximation,
be neglected,  and the integral reduces
to the Fourier transform of the Airy pattern in the object space. Using
\begin{equation}\label{FT}
\iint \e^{i 2\pi( x \xi+ y \eta)} \ \delta_{N}(\xi,\eta)
 \ \d \xi \d \eta=
\mathrm{circ}\left(\frac{x}{N},\frac{y}{N}\right),
\end{equation}
Eq.~(\ref{perf2}) becomes
\begin{eqnarray}\label{perf3}
|U_i(x_i,y_i)|^2 \ &=& \ \frac{1}{|M|^2} \ \left|U_o\left(\frac{x_i}{M},\frac{y_i}{M}\right)\right|^2 \nonumber \\
&\phantom{=}& \  \times \ 
\mathrm{circ}^2\left(\frac{x_i}{Ma},\frac{y_i}{Ma}\right).
\end{eqnarray}
In Eqs.~(\ref{FT}) and (\ref{perf3}),
\begin{equation}\label{disk}
\mathrm{circ}(x,y)=
\cases{%
$1$ & if \ $\sqrt{x^2+y^2}<1,$ \cr 
$0$ & if \ $\sqrt{x^2+y^2}>1.$}
\end{equation}
Eq.~(\ref{perf3}) shows that the intensity in the image plane
exhibits a cut-off. No intensity reaches the image
plane at a distance higher than $Ma$ from the optical axis. This can
be understood in the following way: since the field has been assumed
to be slowly varying, the diffraction in the propagation from the object plane to the
lens is negligible and the limited aperture of the lens has the same
effect as a stop of radius $a$ in the object plane. The function
$\mathrm{circ}\left(x_i/(Ma),y_i/(Ma)\right)$ is the image of that
virtual stop and can be interpreted as the shadow of the lens. This is, however, only a first order approximation,
since the quadratic phase terms $\xi^2/2$ and $\eta^2/2$ in
Eq.~(\ref{perf2}) have been neglected. The effect
of these quadratic phase terms is to create radial intensity
oscillations in the image, especially around the cut-off radius
$Ma$. 

In Fig.~\ref{iso1}, the preceding discussion is illustrated
with an example. A square object
of size $9.5\times 9.5$~mm$^2$ is imaged using a lens of diameter
$2a$ varying from 6~mm to 22~mm. The object can be seen as a plane
screen with a square aperture in it which is illuminated by a plane
wave propagating along the optical axis. The object field
$U_o(x_o,y_o)$ is 1 inside the square and zero outside of it. Fig.~\ref{iso1}
shows the intensity distribution in the image plane
computed using Eq.~(\ref{transPHCni}). With small lenses ($2a$ up to
10~mm) the clipping predicted by Eq.~(\ref{perf3}) is
observed. One can clearly distinguish the disk (\ref{disk}) that
limits the observable part of the object, as well as the intensity ripples
due to the quadratic phase terms in (\ref{perf2}) which were
neglected when deriving (\ref{perf3}). Interestingly, the intensity
oscillations do not disappear as soon as the lens becomes bigger
than the object. For $2a=22$~mm, some residual modulation still
remains. Note that the parameters of this simulation are realistic
ones: the object distance $z_o$ has been fixed to 2~m (close to
infinite-conjugate ratio imaging) and the focal length $f$ to 12~mm
(the numerical aperture ranges from 0.24 to 0.67). Note that energy is lost when imaging large objects through small lenses: For
the simulations shown in Fig.~\ref{iso1}, the percentage of
transmitted energy is, from left to right and top to bottom, 31.1\%,
56.1\%, 83.4\%, 96.6\%, 99.2\%, 99.8\%, and nearly 100\% for the last three images.

The discussion leading to Eqs.~(\ref{perf}) to (\ref{perf3}) only
concerned object fields $U_o(x_o,y_o)$ that are slowly varying on
the length scale $e_o=1.22\ \lambda z_o/a$ in the object plane. For
quickly varying fields the previous discussion
does not hold, but non-isoplanatism still has some effects on
imaging. The way non-isoplanatism modifies the intensity distribution in the
 image plane strongly depends on the object wavefront. No general features can
 be drawn in that case.
To get some insight, consider the following example. 

A large square grid
 ($19\times 19 \ \mathrm{mm}^{2} $) of mutually coherent point sources is imaged through a lens
 of diameter $2a=8$~mm. The imaging conditions are otherwise the same as
 in Fig.~\ref{iso1}. Let's consider that all the point sources are in phase and
 discuss the image formation when the spacing $d$ between the point sources is varied.
\begin{figure}[t]
\centerline{\includegraphics*[width=8.5cm]{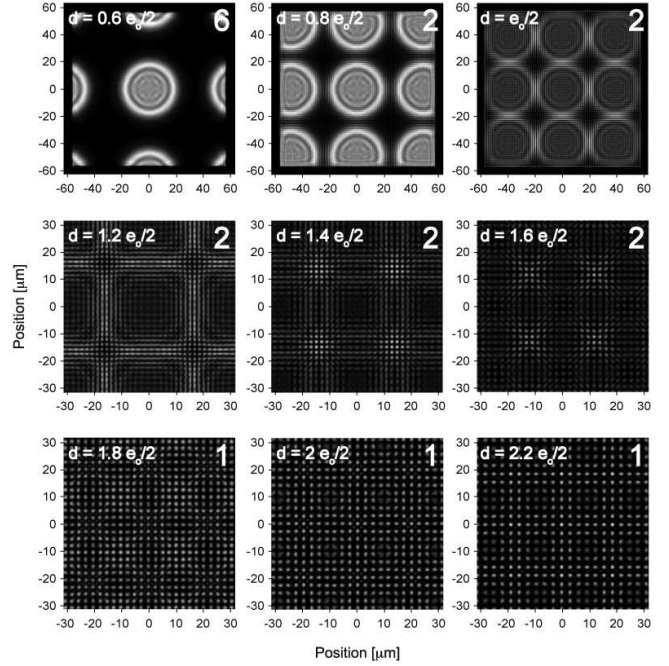}}
\caption{Intensity in the image plane $\mathcal{I}$ when a
large squared ($19\times 19$~mm$^{2}$) grid of point sources, centered on the optical axis, is imaged by a thin
lens. The point sources are mutually coherent and in phase. The different panels correspond to different values of the grid period $d$.  From left to right and top to bottom, $d$ is varied from $0.6 \ e_{o}/2$ to $2.2\ e_{o}/2$ ($e_{o}/2$ is the separation corresponding to the Rayleigh resolution criterion).  
The relevant parameters are: $a=4$~mm, $\lambda=780$~nm, $f=12$~mm, and $z_o=2$~m. The scale of the figures is expressed in microns. The intensity scale is arbitrary but the relative dynamical range is indicated by the number in the upper right corner of each panel. For instance, the number ``6'' in the first panel means that the intensity range is 6 times larger than in the three panels of the last row.} \label{iso2}
\end{figure}
If the points are well separated ($d \gg e_{o}/2$, so that the
Airy patterns associated with them in the image
plane do not overlap) no interference takes place and
non-isoplanatism has no effect on the intensity
distribution in the image plane. If the Airy patterns overlap,
interferences similar to those depicted in Fig.~\ref{imagePHC}
occur. Two cases must be distinguished:  $d\ll e_{o}/2$ and $d\approx e_{o}/2$. For $d\ll e_{o}/2$,  the
 object field varies slowly in space; a fringe pattern similar to
 the one in  Fig.~\ref{iso1} is expected. The interesting case
 is $d\approx e_{o}/2$.  Fig.~\ref{iso2} shows the intensity distribution in
 the image plane when $d$ ranges from 0.6 to 2.2 times $e_o/2$.
 For $d\le e_{o}/2$, a fringe pattern similar to the one in Fig.~\ref{iso1} is
 seen in the center of the field. However, periodic replicas of this
 pattern are also observed. For $d > e_{o}/2$, the circular fringe patterns intersect
 each other, but the sparse sampling due to the grid structure of the image makes
 this structure barely visible (periodicity however remains). For clarity, only the
 central $30 \times 30$~\textmu m$^{2}$ region of the image is displayed in the
 last six panels of Fig.~\ref{iso2}. The simulations of Fig.~\ref{iso2} show that
 the intensity distribution in the image plane exhibit two distinct pseudo-periods 
(in both $x$ and $y$ directions): the small-scale pseudo-period $Md$ due to the grid
structure of the object and the large-scale pseudo-period $X$ associated with the ring
patterns due to non-isoplanatism. Strictly speaking the image is periodic only if $X$
is an integer multiple of $Md$, in which case the period of the image is $X$. Hereafter, we 
use this property to deduce the value of $X$. The object field is modelled as a 
two-dimensionnal Dirac comb:
\begin{displaymath}\label{grid}
U_o(x_o,y_o)= \sum_{n,m} \ \delta(x_o-n \ d,y_o-m \ d),
\end{displaymath}
where the integers $n$ and $m$ run from
$-\infty$ to $+\infty$. Using Eq.~(\ref{transPHCni}), we then find that 
\begin{eqnarray}\label{imgrid}
|U_i(x_i,y_i)|^2& =&\frac{1}{|M|^2} \times \nonumber \\
&& \left|\sum_{n,m} \e^{i \phi_{nm} }
\delta_{\frac{a}{\lambda z_o}}(\frac{x_{i}}{M}- n
d,\frac{y_{i}}{M}-m d)\right|^2,
\end{eqnarray}
with
\begin{displaymath}\label{phinm}
\phi_{nm}=\frac{\pi}{\lambda z_o}\left(n^2+m^2\right) d^2.
\end{displaymath}
In order to find $X$, we require that $|U_i(x_i,y_i)|^2=|U_i(x_i+X,y_i)|^2$ when $X$ is a multiple of $Md$.
 Using eq.~(\ref{imgrid}), we have
\begin{eqnarray*}
|U_i&&(x_i+X,y_i)|^2 =\frac{1}{|M|^2} \times \\
&&\left|\sum_{n,m} \e^{i \phi_{nm} }
\delta_{\frac{a}{\lambda z_o}}(\frac{x_{i}+X}{M}- n
d,\frac{y_{i}}{M}-m d)\right|^2.
\end{eqnarray*}
The argument  $(x_{i}+X)/M- n d$, can
be rewritten as $x_{i}/M- n'd$ with $n'=n -X/(Md)\in \mathbb{Z}$.
Replacing the sum over $n$ by a sum over
the values of $n'$, we obtain
\begin{eqnarray*}\label{imgridtrans}
|U_i&&(x_i+X,y_i)|^2 =\frac{1}{|M|^2} \times \nonumber \\
&&\left|\sum_{n',m} \e^{i \phi_{n'm} }
\delta_{\frac{a}{\lambda z_o}}(\frac{x_{i}}{M}- n'
d,\frac{y_{i}}{M}-m d)\right|^2.
\end{eqnarray*}
where
\begin{displaymath}\label{phinpm}
\phi_{n'm}=\frac{\pi}{\lambda
z_o}\left[n'^2+2n'\frac{X}{Md}+m^2\right] d^2.
\end{displaymath}
One can note that  $\phi_{n'm}$ and $\phi_{nm}$ just differ by a factor $n'\times 2\pi$,  and therefore
$|U_i(x_i,y_i)|^2=|U_i(x_i+X,y_i)|^2$, if $X$ is given by
\begin{equation}\label{per}
X=M\times \frac{z_{o}\lambda}{d}.
\end{equation}
When $X$ is not a multiple of $Md$, formula (\ref{per}) is still valid 
if $X$ is understood as the large-scale pseudo-period. However, the notion of ``large-scale pseudo-periode'' 
only holds if $X/(Md)\approx a \ e_o/d^2 \gg 1$  ($e_{o}=1.22 \ \lambda z_{o}/a$).  
Since $a/d$ is usually a large number, this condition can still be valid
 even if $d$ strongly exceeds  the distance $e_{o}/2$ corresponding to the Rayleigh criterion. The
 periodic patterns displayed in Fig.~\ref{iso2} prove that
 the field in the image plane corresponding to the Airy pattern of a given point in the grid
influences the entire image on a length scale much longer than the usually
 considered ``Airy pattern diameter'' $e_{o}$. Readers familiar with signal processing 
will notice that the reason why periodicity appears here is the same one
 that makes the Fourier spectrum of a sampled signal (Dirac comb) periodic. We
however  stress that the periodicity that is described here appears in the image plane
 \emph{and not} in the Fourier plane of the object field. This is a peculiarity of non-isoplanatic imaging.

\begin{figure}[t]
\centerline{\includegraphics*[width=8.5cm]{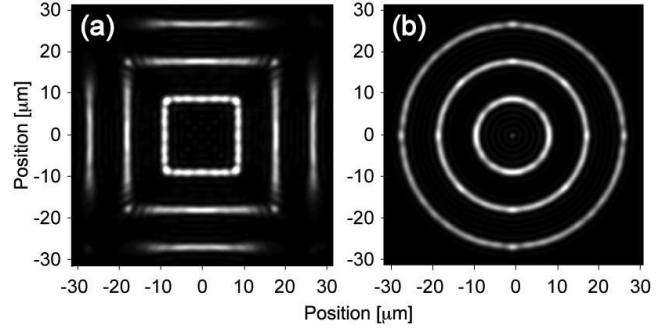}}
\caption{Intensity in the image plane $\mathcal{I}$ when squared and circular contours are imaged by a thin lens. In the object plane, all the point sources forming the contours are mutually coherent and in phase. The imaging conditions are the same as in Fig.~\ref{iso1} ($\lambda=780$~nm, $f=12$~mm, and $z_o=2$~m), but the size of the lens is fixed: $2a=6$~mm, as in the first pannel of Fig.~\ref{iso1}. The intensity scale is arbitrary. Panel (a) shows the image of three squared contours of increasing side $b=a$, $b=2a$ and $b=9.5$~mm~$\approx 3.2 \ a$. The size of the outer square is the same as in Fig.~\ref{iso1}. Panel (b) shows the image of three concentric circles, the diameters of which are equal to the sides of the squares in the panel (a).} \label{iso4}
\end{figure}

Though non-isoplanatism can strongly influence the image formation of both slowly varying and 
rapidly varying object fields, the effects are qualitatively different. For slowly varying fields, 
we have seen that the major effect comes from the shadow that the lens projects in the image plane.
The image is clipped so that only the central disk of radius $Ma$ remains (see Fig.~\ref{iso1}). 
The simulations of Fig.~\ref{iso2} show that, for rapidely varying
object fields, light \emph{is} observed outside the central disk of radius $Ma$. 
More complex effects are exhibited, but clipping do not occur anymore. In many applications of 
coherent optics (like optical lithography), the object is made of lines instead of single points 
or filled surfaces. Lines are objects on which the field varies slowly in one direction (the direction tangeant to the line) and rapidly in the orthogonal one. Therefore, one can expect that the effects of non-isoplanatism will be intermediate between the two previous cases. Fig.~\ref{iso4} helps to understand how lines are imaged through a non-isolanatic optical system. The imaging conditions are the same as in Fig.~\ref{iso1}, except that the lens diameter is fixed: $2a=6$~mm. Panel (a) shows the image of three squared contours. The outer square is exactly the contour of the filled square imaged in Fig.~\ref{iso1}. The comparison with the upper-left pannel of Fig.~\ref{iso1} shows that the major part of the contour is now visible; only the corners of the square are clipped. Closer examination shows that, for a straight line, only a segment of length $2a$ in the object plane ($2 M a $ in the image plane) is transmitted. The part of the line that is clipped corresponds to the orthogonal projection of the shadow of the lens on the straight line. This can be easily understood by analysing the propagation of the cylindrical waves emitted by straight lines through the spherical lens. The smaller square in pannel (a) is transmitted because its side is shorter than the diameter of the lens. The intermediate square is at the limit of the cut-off. As shown in the pannel (b) of Fig.~\ref{iso4}, circles are never clipped, whatever their radii, because the orthogonal projection of the lens disk on the cicle is the circle itself.

\section{Isoplanatic imaging through a thin
lens}\label{sec4}
 In Sec.~\ref{sec3}, we claimed that
isoplanatic imaging is possible with a thin lens
when the object lies on the spherical surface $\mathcal{S}_o$ and
the image is observed on the spherical surface $\mathcal{S}_i$.
Let's examine this statement more closely.

Referring to Fig.~\ref{tl}, simple Gaussian Optics arguments suggest
that if the point $P_2$ is translated horizontally from the object plane $\mathcal{O}$ to the surface
$\mathcal{S}_o$, its image $P'_2$, instead of moving
towards $\mathcal{S}_i$, should move away from the lens.
This argument is correct, but it relies on the Gaussian approximation
that the point $P'_2$ \emph{is} initially in the ``image plane''
$\mathcal{I}$.  In reality, because of the field
curvature aberration due to the lens, the
stigmatic image of $P_2$ is closer to the lens than the surface
$\mathcal{S}_i$ itself ($P'_2$ lies on the so-called Petzval surface). Bringing $P_2$ on
$\mathcal{S}_o$ will place its image $P'_2$ exactly on
$\mathcal{S}_i$ \cite{Hecht}.
Consequently, Eq.~(\ref{transPHCi}) is exact,
while Eq.~(\ref{transPHCni}) is only valid in the context of
Gaussian approximation. If the lens is still diffraction-limited in
this regime (no point-aberrations), the isoplanatic imaging geometry
provides a nearly perfect transfer of the coherent field from the
object to the image space. Only the resolution is reduced
because of the finite size of the lens.

The problem that remains is how to transfer
a given field from a plane surface to a curved one
before imaging, and vice versa after imaging.
Generally, the object field is obtained by
modulating a coherent illumination field $A(x_o,y_o)$ by a complex
amplitude function $u_o(x_o,y_o)$:
\begin{equation}
U_o(x_o,y_o)=A(x_o,y_o) \times u_o(x_o,y_o).
\end{equation}
The modulation $u_o(x_o,y_o)$ can be produced by any plane
modulation device like a plane transparency, a grating or some kind of
adaptative optics. Usually $u_o(x_o,y_o)$ itself is considered as
the ``object'' of interest. In that case, the illumination beam must
be a plane wave propagating along the optical axis in order to map
the profile $u_o(x_o,y_o)$ to the beam: $U_o(x_o,y_o)\propto
u_o(x_o,y_o)$. If, instead, we illuminate the modulation device with
a spherical wave $A(x_o,y_o)=A\exp{(-i 2 \pi r/\lambda)}$ focusing
on $O$ (see Fig.~\ref{tl}), we get
$U_{\mathcal{S}_o}(x_o,y_o)=U_{o}(x_o,y_o)\exp{(i 2 \pi
r/\lambda)}\propto u_o(x_o,y_o)$. This kind of illumination projects
the object onto $S_o$ as required for isoplanatic imaging
\cite{Good}. Starting with an illuminating plane
wave travelling along the optical axis, this can
be achieved by placing an additional thin lens of
focal length $z_o$ just before (or just after) the object plane $\mathcal{O}$. This
lens must act as a pure phase-correction transparency.
Similarly, on the image side, a thin lens of focal length $z_i$
placed just after (or before) the image plane can be used to
project the image from $S_i$ onto $\mathcal{I}$.
From a broader point of view, any spherical field distribution in
the object space of a centered paraxial optical system can be imaged
onto a spherical surface of any curvature using only lenses;
some of them will be imaging lenses, while others
will play the role of phase-correction transparencies. This is the
basis of the so-called \emph{metaxial optics} theory formulated by
the Bonnet
\cite{bonnet1,bonnet3,bonnet2}.

It should be noted that this approach only works
for sufficiently slowly varying fields,
because the diffraction from $\mathcal{O}$ to $\mathcal{S}_o$ (and
$\mathcal{S}_i$ to $\mathcal{I}$) has to be
negligible for the amplitude of
the fields on $\mathcal{O}$ and $\mathcal{S}_o$
($\mathcal{S}_i$ and $\mathcal{I}$)
being the same. In addition, due to optical design
constraints, the use of optical lenses as phase-correction
transparencies is not always possible. Moreover,
lenses are never perfectly thin and diffraction through them may
have worse effects on imaging than the phase
curvature due to non-isoplanatism. Whether or not non-isoplanatism
should be corrected depends on the particular system under
consideration. A clever design can minimize, if not cancel, its
effects on the detected intensity.

\section{Experimental investigation}
The theoretical discussion of Secs.~\ref{sec5} and \ref{sec4}
relies on two strong approximations: the paraxial approximation and
the thin lens approximation. One may wonder whether our analysis is
robust enough to be applied to systems containing powerful lenses,
which are usually thick and have a high numerical aperture. The
following experiment shows that the previous discussion is also valid for these systems.

\begin{figure}[b]
\centerline{\includegraphics*[width=8cm]{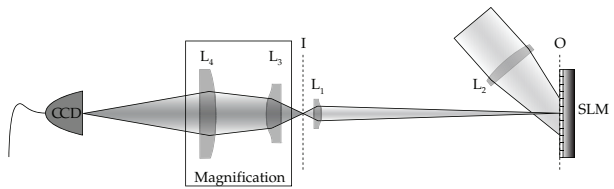}}
\caption{Experimental setup: L$_1$, aspheric lens ($f=8$~mm);
L$_2$, achromatic doublet ($f=750$~mm); L$_3$, aspheric lens ($f=20$~mm); and L$_4$, achromatic
doublet ($f=500$~mm); SLM, spatial light modulator; CCD, coupled-charge
camera; O, object plane; I, image plane.} \label{setup}
\end{figure}
\begin{figure}[p]
\centerline{\includegraphics*[width=8.5cm]{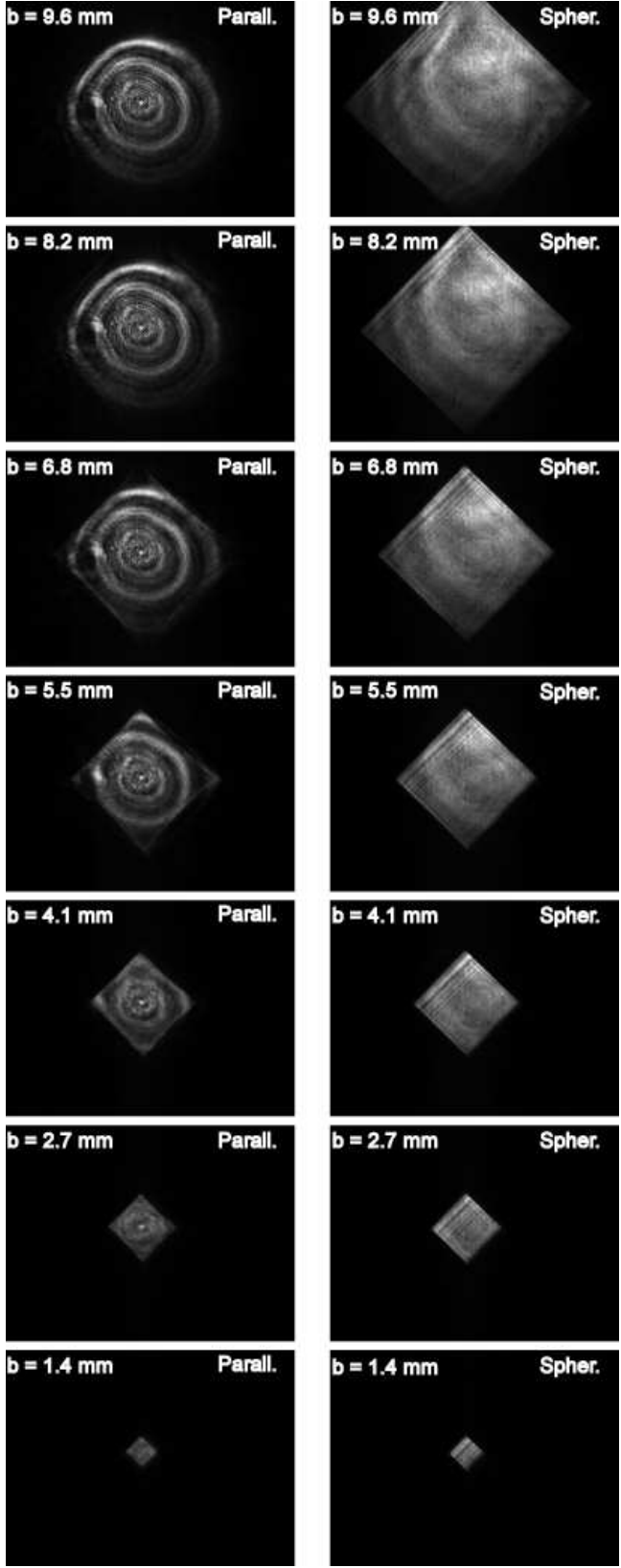}}
\caption{Images of different-sized squares (the side $b$ ranges from
1.4~mm to 9.6~mm) recorded using the setup in Fig.~\ref{setup}. The
object plane is either illuminated with a parallel beam
(non-isoplanatic case, left column) or with a spherical one
converging on L$_1$ (isoplanatic case, right column).}
\label{expSQ}
\end{figure}

The setup is shown in Fig.~\ref{setup}. The test lens L$_1$ is an aspheric lens
 having a focal length $f=8$~mm and a
diameter $2a=8$~mm (LightPath 352240). This lens is used to image the
surface of a spatial light modulator (SLM), a $1024 \times 768$ micromirror array, that modulates the amplitude of the reflected beam. Using
 the SLM, we can generate arbitrary patterns. The resolution is set by the size
 of the micromirrors ($13 \times 13$~\textmu m$^{2}$). The object plane is 60~cm away from
 the lens. We use a double-lens
system (L$_3$, L$_4$) to magnify the image produced and project it
on the CCD camera.  L$_3$ is a
diffraction-limited aspheric lens and L$_4$ a long-focal achromatic doublet. The numerical aperture
 of this double-lens system is large enough to prevent any possible clipping
 or diffraction during the magnification process. The SLM is
either illuminated with a plane wave or a spherical wave converging
on L$_1$ ($\lambda=780$~nm). In the first case, the imaging system
is non-isoplanatic. As explained in Sec.~\ref{sec4}, it becomes
isoplanatic when a spherical-wave illumination is used. The
spherical wave is obtained from the impinging plane wave  by
inserting the additional lens L$_2$ (75-cm focal length, achromatic doublet)  in
front of the SLM. We use this setup to image objects of different sizes
and shapes through the test lens.

Fig.~\ref{expSQ} shows the pictures recorded by the CCD camera when
we image squares of different sizes through the 8-mm diameter
aspheric lens L$_1$. Let's first consider the case when the SLM 
is illuminated by a parallel beam (non-isoplanatic imaging, 
left column of Fig.~\ref{expSQ}). For a square side larger than 8~mm, 
the clipping effect described in Sec.~\ref{sec5} is clearly observed. 
 Circular fringes similar to those of Fig.~\ref{iso1} are also seen
 in the image plane. 
For a square side smaller that 8~mm, the region of non-zero intensity is 
limited by the size of the square. However, intensity modulation due to 
non-isoplanatism is still noticeable for a square side as small as 2.7~mm. 
In the case of the $1.4\times1.4$~mm$^{2}$ square, non-isoplanatism has a negligible 
effect. Note that, in the simulations of Fig.~\ref{iso1}, the lens size was 
varied while the object size was kept constant. Here, the lens is always the 
same, but the square size is varied instead. For this reason, the ring pattern 
is the same for every picture in the left column of Fig.~\ref{expSQ}. When the 
SLM is illuminated with a spherical wave converging on L$_{1}$ (isoplanatic imaging, 
right colomn of Fig.~\ref{expSQ}) no clipping effect occurs and there is, in 
principle, no limit to the size of the objects that the system can image (the slight 
variations in intensity are due to amplitude inhomogeneities is the illumination beam).

\begin{figure}[t]
\centerline{\includegraphics*[width=8cm]{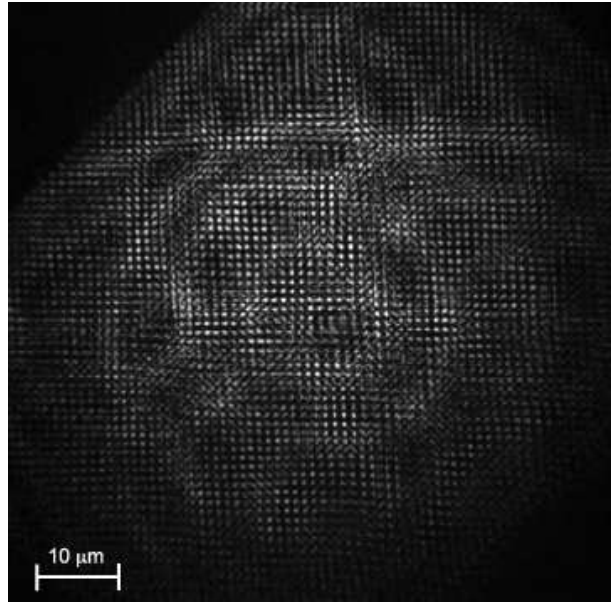}}
\caption{Images of a rectangular grid of points recorded using the setup in Fig.~\ref{setup}. The
object plane is  illuminated with a parallel beam
(non-isoplanatic case). The period of the grid in the object plane, $d=109$~\textmu m, corresponds to 1.5 times the Rayleigh criterion separation ($e_{o}/2$).}
\label{Expgrid}
\end{figure}

In a second experiment, we imaged rectangular grids of points illuminated by a plane wave (non-isoplanatic illumination), 
a situation that we described theoretically in Sec.~\ref{sec5}. 
Fig.~\ref{Expgrid} shows
 the recorded intensity distribution in the image plane for a grid with a 
period $d=109$~\textmu m, which corresponds to 1.5 times the Rayleigh criterion
 separation ($e_{o}/2$). As shown in Sec.~\ref{sec5}, for such large value of $d$ the
 ring patterns should not be visible anymore (see the six last panels of Fig.~\ref{iso2}). Here,
 however, we can distinguish one central ring pattern and four replicas intersecting it. We
 attribute this discrepancy to the fact that the experimental point-spread function is
 a bit broader than the theoretical Airy pattern. As noted is Sec.~\ref{sec5}, the
 ``disappearance'' of the multiple ring patterns when $d$ increases is caused by the sparseness
 of sampling due to the grid. A broadening of the point-spread function reduces that sparseness
 and therefore restores the ring patterns characteristic of non-isoplanatic imaging.

 \section{Conclusion}
In general, even diffraction-limited imaging systems distort the phase of the processed fields.
 This is of no relevence when used with incoherent light, but has a tremendous effect on coherent 
imaging. In combination with Fraunhofer diffraction from the finite instrumental aperture, the phase
 distortion leads to a severe degradation of the field amplitude in the
 image plane. We analyzed this phenomenon for two very different but prototypic
 imaging systems (the pinhole camera and a thin lens) and observed its
 fundamental and general nature. We showed that substantially different effects arise depending on 
 whether the field varies slowly
or rapidly on the length scale of an Airy pattern. The degradation of the field
 amplitude can however be overcome or, at least, limited by a clever design
 of the optical system. The main aspects
of our analysis have been confirmed experimentally using a 
powerful thick aspheric lens to demonstrate that the phenomenology that is
described also holds beyond the paraxial and thin lens approximations
used for the theoretical analysis.

\section*{Acknowledgments}

We gratefully acknowledge support by the Engineering and Physical Sciences Research Council (EP/E023568/1), the QIP IRC (GR/S82176/01),
the Research Unit 635 of the German Research Foundation, and the EU through the research and training network EMALI 
(MRTN-CT-2006-035369) and the integrated project SCALA. One of us (E. B.) also wishes to express his gratitude to the Philippe Wiener
and Maurice Anspach Foundation.


%






\end{document}